# DECODING AI AND HUMAN AUTHORSHIP: NUANCES REVEALED THROUGH NLP AND STATISTICAL ANALYSIS

Mayowa Akinwande[1], Oluwaseyi Adeliyi[2], Toyyibat Yussuph[3]

[1]Department of Computer Sciences, Austin Peay State University, USA,
[2] University of Arkansas at Little Rock, USA
[3]American Express Company, Arizona, USA

## ABSTRACT

*This research explores the nuanced differences in texts produced by AI and those written by humans, aiming to elucidate how language is expressed differently by AI and humans. Through comprehensive statistical data analysis, the study investigates various linguistic traits, patterns of creativity, and potential biases inherent in human-written and AI-generated texts. The significance of this research lies in its contribution to understanding AI's creative capabilities and its impact on literature, communication, and societal frameworks. By examining a meticulously curated dataset comprising 500K essays spanning diverse topics and genres, generated by LLMs, or written by humans, the study uncovers the deeper layers of linguistic expression and provides insights into the cognitive processes underlying both AI and human-driven textual compositions. The analysis revealed that human-authored essays tend to have a higher total word count on average than AI-generated essays but have a shorter average word length compared to AI-generated essays, and while both groups exhibit high levels of fluency, the vocabulary diversity of Human authored content is higher than AI generated content. However, AI-generated essays show a slightly higher level of novelty, suggesting the potential for generating more original content through AI systems. The study also identifies a lower prevalence of gender bias in AI-generated texts but a higher presence of biased topics overall. These findings highlight the strengths and limitations of AI in text generation and the importance of considering multiple approaches for comprehensive analysis. The paper addresses challenges in assessing the language generation capabilities of AI models and emphasizes the importance of datasets that reflect the complexities of human-AI collaborative writing. Through systematic preprocessing and rigorous statistical analysis, this study offers valuable insights into the evolving landscape of AI-generated content and informs future developments in natural language processing (NLP).*

## 1. INTRODUCTION

In recent years, the integration of Artificial Intelligence (AI) into various aspects of our lives has brought about significant changes in content creation. With the emergence of AI-driven language models, particularly generative text algorithms, the lines between human-authored texts and machine-generated content have become increasingly blurred. Despite widespread assumptions, often based on anecdotal evidence, suggesting that AI could replace human involvement in writing activities [1], this research project seeks to discover and understand the subtle differences that distinguish AI-generated language from human-authored writing. The primary goal of this research is to explore how AI and humans' express language differently. By examining various linguistic traits, patterns of creativity, and potential biases inherent in AI-generated and human-authored texts [2], the aim is to shed light on AI's capabilities in content production.

This investigation goes beyond surface-level analysis, aiming to uncover the deeper layers of linguistic expression and provide insights into the creative features and cognitive processes underlying both AI and human-driven textual compositions. The significance of this paper lies in its contribution to the ongoing discourse surrounding AI's creative capabilities and its impact on diverse fields such as literature, communication, and knowledge dissemination. By revealing the intricate complexities present in generative AI, it aims to facilitate informed discussions about responsible AI usage and the ethical considerations of incorporating machine-generated content into societal frameworks [3]. Additionally, the study addresses the critical need to uncover potential biases or disparities in thematic substance between AI and human writing, not only to improve AI algorithms but also to ensure that AI-generated content adheres to ethical norms and cultural sensitivities.

The dataset utilized in this research was sourced from ShaneGerami's AI vs Human Text on Kaggle [4], it comprises of 500K essays with two columns: "text" and "generated" (AI=1, Human=0). The focus is on analyzing the linguistic features, creativity metrics, and potential biases present in the essays. The textual data will be preprocessed, and statistical analyses will be conducted to quantify differences between AI and human writing. The uniqueness of this dataset lies in its meticulous curation, ensuring a balanced representation of both AI-generated and human-written compositions. By including essays from different sources and annotators, the dataset captures a wide spectrum of writing styles and complexities inherent in language generation tasks. This breadth and depth enable thorough analysis and exploration of the nuances in language expression between AI and human authors. Through systematic preprocessing and rigorous statistical analysis, this dataset serves as a cornerstone for unraveling the intricatedynamics of generative AI and human-authored texts, contributing valuable insights to the research landscape. As the research navigates through the statistical analysis of these large datasets comprising both AI and human-generated texts, it aims to offer valuable insights into the evolving landscape of AI-generated content. One of the challenges faced in assessing the language generation capabilities of large language models (LMs) within NLP is that datasets often overlook the intricacies of the writing process, prioritizing final outcomes over interactive collaboration. To address this challenge, the approach involves designing reusable and expandable datasets that capture the nuances of the writing process. This approach facilitates the consideration of interactive settings in LLM evaluation and highlights the necessity for datasets that reflect the complexities of human-AI collaborative writing [5]. The outcomes of this exploration are expected to foster a deeper understanding of the creative capabilities of AI, thereby informing future developments in natural language processing and contributing to the responsible advancement of AI technologies.

## 2. Literature Review

Linguistic analysis, a fundamental element of NLP research, is explored in Marjorie McShane and Sergei Nirenburg book "Linguistics for the Age of AI," which delineate key tenets guiding linguistic work in the context of AI integration [6]. The aim of artificial intelligence (AI) has been to develop intelligent systems capable of using language as proficiently as humans, facilitating fluent conversations and a nuanced comprehension of language intricacies. According to McShane and Nirenburg, language processing within AI models is conceptualized from an agent perspective, integrated into a broader model of perception, reasoning, and action. Central to this perspective are the core prerequisites for success, including the ability to extract meaning from linguistic expressions, represent them in memory, and utilize these representations for decision-making across verbal, physical, and mental actions. The multifaceted nature of linguistic phenomena ranges from morphological ambiguity to pragmatic ambiguity. Semantic analysis emerges as a pivotal sub-task of NLP, enabling computers to derive meaning from textual data through grammatical analysis and contextual interpretation.

Semantic classification models, including topic classification, sentiment analysis, and intent classification, demonstrates the practical applications of semantic analysis in various domains, from customer service to marketing analytics. Linguistic analysis provides a rich theoretical framework and methodological insights crucial for understanding the complexities of language generation in AI systems. By leveraging the theoretical foundations and practical methodologies outlined in the research, it contributes to the understanding of linguistic phenomena in both AI-generated and human-authored texts, and highlights both successes and limitations in statistical methods. Syntactic parsing, semantic role labeling, coreference resolution, and distributional semantics have demonstrated utility in NLP tasks, such as question answering and information extraction. However, challenges persist in handling semantic ambiguity and achieving consistent performance across different linguistic phenomena. The ongoing refinement of statistical models and the integration of linguistic features inform ongoing efforts to enhance language understanding systems and improve their performance across diverse tasks. The convergence of Big Data and Artificial Intelligence has propelled the field of text analytics into the spotlight, offering profound insights into linguistic structures and patterns inherent in vast volumes of textual data [7]. Antonio Moreno and Teófilo Redondo explore the interdisciplinary realm of text analytics, shedding light on its significance and applications, which extend to linguistic analysis. At its core, text analytics encompasses the extraction of valuable linguistic information from diverse textual sources, including emails, blogs, tweets, and forums. This process, often referred to as text mining, falls within the broader scope of NLP, a foundational branch of Artificial Intelligence dating back to the 1950s. While text analytics primarily focuses on uncovering new insights and knowledge from written resources, its applications extend beyond linguistic analysis to encompass various domains such as sentiment analysis, customer feedback analysis, and topic tracking. Within the domain of NLP, text analytics serves as a vital tool for understanding the intricate nuances of language usage, including syntactic structures, semantic meanings, and pragmatic implications. Techniques such as information extraction, topic tracking, summarization, and categorization enable researchers to delve deep into linguistic phenomena and extract actionable insights from textual data. However, the challenge of understanding figurative language, including irony and metaphor, remains a persistent obstacle, requiring contextual interpretation and domain-specific knowledge. Despite these challenges, the integration of text analytics with linguistic analysis holds immense promise for advancing our understanding of language and communication. By leveraging AI-driven approaches and Big Data analytics, researchers can navigate the complexities of linguistic structures, uncover hidden patterns, and unlock new realms of knowledge embedded within textual data. As organizations increasingly rely on textual data for decision-making and strategic insights, the role of text analytics in linguistic analysis becomes ever more crucial in harnessing the power of language for actionable intelligence.

The emergence of generative artificial intelligence (AI) challenges traditional notions of creativity, prompting a reevaluation of its essence and its relationship to human creativity [7]. Generative AI exhibits an uncanny ability to produce original content resembling human creative choices, such as writing, painting, and composing music, blurring the lines between human and machine creativity. Despite operating on algorithmic principles, generative AI derives its rules from training data, simulating human-like creative processes. Two responses have emerged in the creative sector: one suggesting that AI lacks individual expression characteristic of human creativity, while the other argues that AI merely recombines existing cultural elements into new forms, devoid of genuine creativity. The rise of generative AI challenges conventional notions of creativity, raising fundamental questions about its nature and the role of machines in creative endeavors. To contextualize the current research within the broader landscape of existing studies, a comparative analysis of related work is presented in Table 1. This table summarizes the focus, techniques used, key findings, and distinctions between the proposed research and existing studies.

Table 1. Comparative Analysis of Related Work

| Study | Focus | Techniques Used | Key Findings |
|---|---|---|---|
| Tony Berber Sardinha (2024) | AI-generated vs Human-authored texts | Multidimensional comparison | Differences in linguistic patterns |
| Yongqiang Ma et al. (2023) | Differentiation Analysis of Scientific Content | Analysis of scientific text generation | AI's influence on scientific writing |
| Elisha Nañola et al. (2023) | Comparative Voice Analysis of Academic Essays | Voice Analysis | Differences in AI and student-authored academic essays |
| McShane &Nirenburg | Linguistics for AI | Agent-based language processing | Linguistic work in AI Integration |
| Bommasani et al (2023) | Holistic Evaluation of Language Models | Comprehensive benchmarking | Holistic view of language model capabilities |

## 3. Natural Language Processing (NLP) Techniques

Drawing on theories of language generation, cognitive linguistics, and computational creativity to establish a foundation for this study we explore a dataset which comprises of 500K essays with two columns: "text" and "generated" (AI=1, Human=0). The focus is on analyzing the linguistic features, creativity metrics, and potential biases present in the essays. Textual data is preprocessed, and statistical analyses is conducted to quantify differences between AI and human writing. Key NLP techniques used on the dataset are further discussed below

### 3.1. Tokenization and Part-of-Speech Tagging

The texts are broken down into individual tokens (words or sub words). This is essential for subsequent analyses and assigned grammatical categories (e.g., noun, verb, adjective) to each token. This helps in understanding the syntactic structure of sentences. After loading the Original dataset size: (487235, 2) and defining a subset size of 100000, which is a random subset of the data, a copy of the Data Frame is created and then the raw texts are tokenized into words or sub words for the random subset, the list of tokens is converted to a single string for each row in the 'text' column while exploratory data analysis on the data subset gives the distribution of essay lengths Fig. 1. Following tokenization, word frequency was calculated, and the most common words were visualized. This analysis yielded insights into the vocabulary size, which was found to be 121,790, and the total number of words in the essays, totaling 43,976,390.

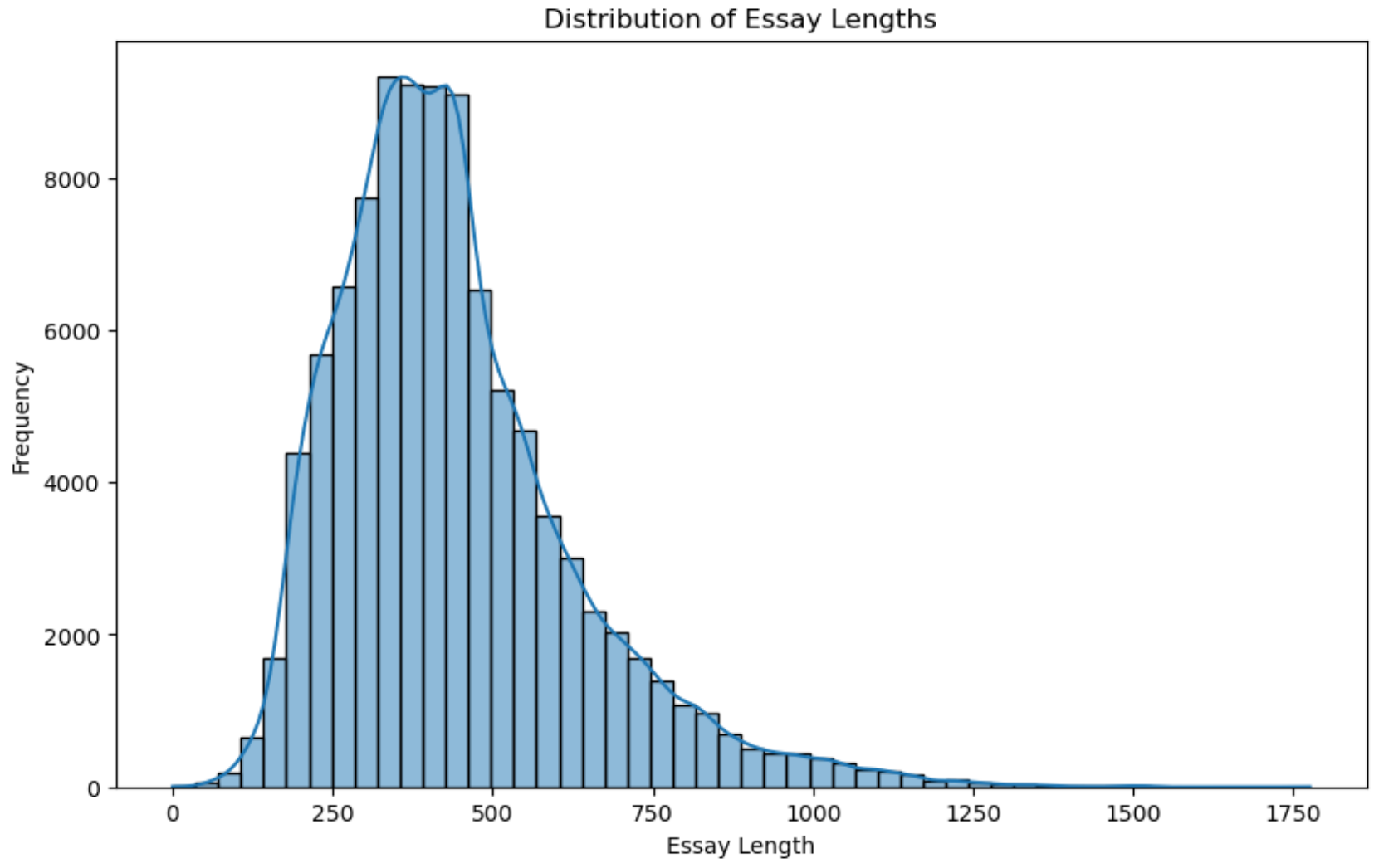

Fig.1. The distribution of essay lengths

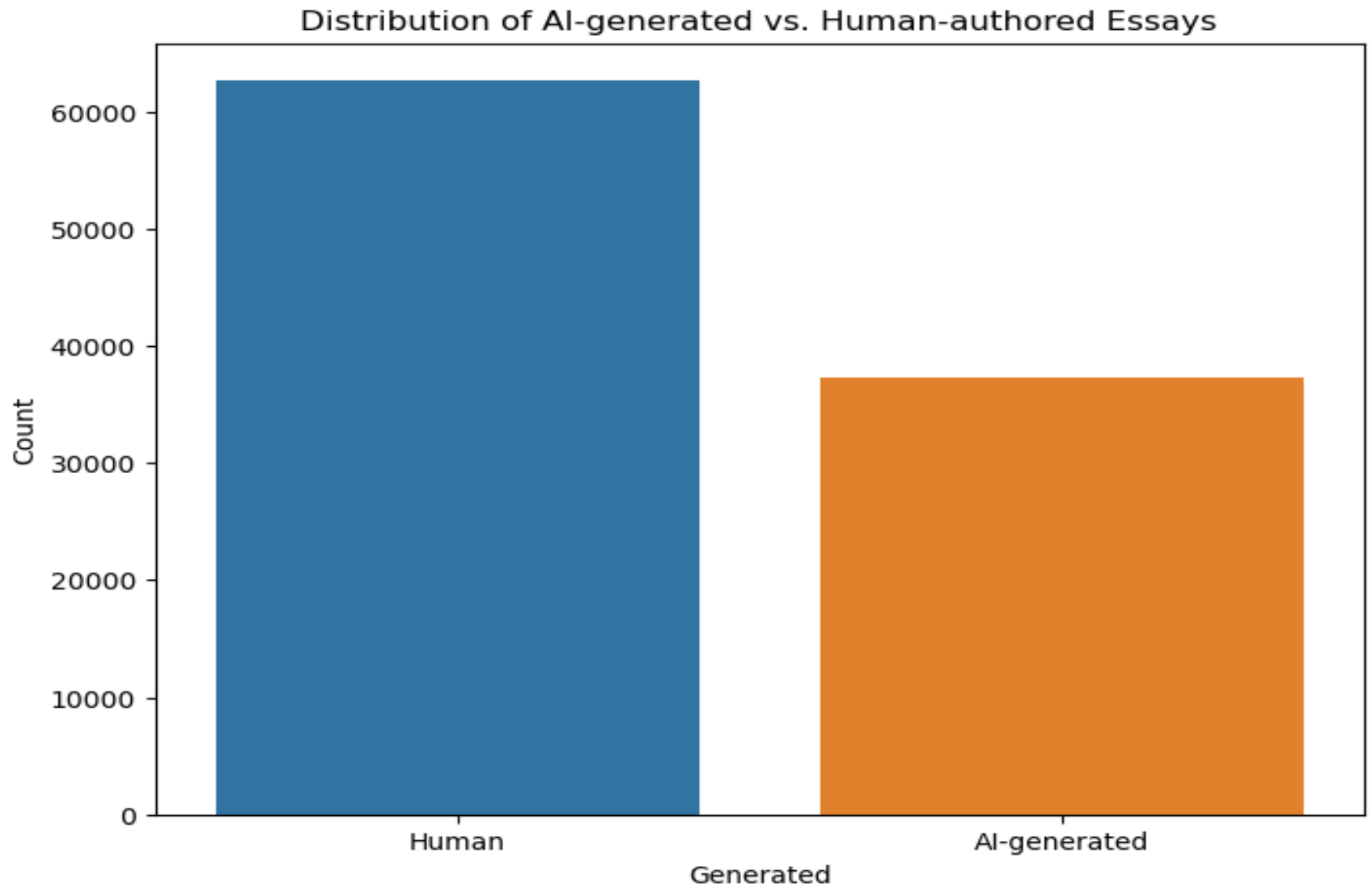

Fig.2. The distribution of AI-generated versus human-authored essays

A count plot is created to visualize the distribution of AI-generated versus human-authored essays within the dataset. The 'generated' column was used as the categorical variable, with values representing whether an essay was authored by a human (labeled as 0) or generated by an AI (labeled as 1). Fig 2 displays the frequency of each category, offering a clear comparison between human and AI-authored essays.

## 3.2. Sentiment Analysis

Sentiment analysis is a way to sort texts that focuses on figuring out what subjective words mean with aim to discern public sentiment by analyzing opinions [9]. To determine the sentiment expressed in each essay (positive, negative, neutral). This can reveal emotional tones in both AI and human-generated texts. Sentiment analysis is pivotal in the realm of polarity detection and emotion recognition, targeting entities ranging from individuals to topics and events. Its significance transcends various domains, finding extensive utility in both business and social networks across diverse applications [10]. By dissecting textual data and discerning the prevailing sentiments, it enables nuanced insights into user perceptions, attitudes, and reactions. This deeper understanding fosters more informed decision-making processes and facilitates the identification of trends and patterns within the data. Sentiment analysis, also known as opinion analysis or opinion mining, has been explored across various levels: Document Level, Sentence Level, Phrase Level, and Aspect Level [11]. In this paper, we analyze the sentiment expressed within entire documents or essays, rather than focusing solely on the sentence, phrase, or aspect level.

### 3.3. Theme Analysis

Thematic analysis involves an emergent and interactive process of interpretation applied to a collection of messages, typically resulting in thematic structure. It entails identifying and analyzing recurring themes within textual or qualitative data, which illuminate underlying ideas, concepts, or patterns, thereby offering insight into the data's deeper meaning or message. Additionally, the process aims to identify and classify entities, such as person names, locations, and organizations, within the text. By employing techniques like Named Entity Recognition using spaCy, it becomes possible to discern discrepancies in the types of entities mentioned by AI-generated content compared to those authored by humans.

### 3.4. Lemmatization

Lemmatization plays a crucial role in standardizing words to their base or root form. By doing so, it facilitates the recognition of common language patterns and ensures that variations of words are treated consistently throughout the analysis. This process enhances the accuracy and depth of our linguistic exploration by capturing the essence of words and their semantic connections. Lemmatization is the process of grouping the various inflected forms of a word to recognize them as a unified entity, referred to as the word's lemma or its vocabulary form [12]. While similar to stemming, lemmatization goes a step further by preserving the semantic meaning of individual words. In essence, it consolidates text containing similar meanings into a single word. lemmatization employs an algorithmic technique to determine the lemma of a word, which represents its root form rather than merely its stem.

### 3.5. Text Vectorization

Text vectorization involves converting textual data into numerical vectors, enabling the application of machine learning algorithms. Using TF-IDF and word embeddings on the dataset.

$$TF - IDF(t, d, D) = TF(t, d) \times IDF(t, D)$$

Where TF(t,d) is the term frequency of term t in document d and IDF(t,d,D) is the inverse document frequency of term t in document set D. This process is essential because raw text data cannot be directly utilized for model parameter training; thus, it must be transformed into numerical format for feature extraction. Text vectorization can be achieved through two primary methods: word vectorization and paragraph vectorization [13]. In this research word vectorization is implemented where each word in the text is represented as a numerical vector and as distributed representations, capturing both syntactic and semantic information. Before inputting text data into neural network layers, it must be vectorized using a suitable method to convert it into structured data. After text is turned into word vectors, discrete symbols are stored based on their indices to lower the number of parameters and make generalization easier.

### 3.6. Syntax and Dependency Parsing

To analyze the grammatical structure of sentences, including relationships between words, advanced techniques such as deep-syntactic dependency structures is used to capture intricate relationships within sentences, including argumentative, attributive, and coordinative relations among words. These structures offer significant potential for numerous NLP applications by providing a nuanced understanding of sentence composition. Syntax and dependency parsing are integral components of natural language processing, aimed at analyzing the grammatical structure of sentences and elucidating the relationships between words [14]. This analysis provides insights into sentence complexity and structure, aiding in various NLP tasks. Hereby exploring the nuanced interplay between syntax and semantics in natural language processing, utilizing sophisticated parsing techniques to extract rich linguistic information from textual data. Dependency syntax has proven to be immensely valuable for various NLP tasks, including those relevant to this research [15]. Common methodologies for leveraging dependency syntax include Tree-RNN and Tree-Linearization, both of which utilize explicit 1-best tree outputs from proficient parsers as inputs. These techniques potentially enhance the understanding and analysis of the grammatical structure of text data.

### 3.7. Language Model Evaluation

Language model evaluation serves as a cornerstone for comprehending AI systems, providing valuable insights into their capabilities and limitations. By developing comprehensive benchmarks that cover various aspects of LM performance, we can gain a nuanced understanding of their functionality. These benchmarks not only guide the refinement of existing LMs but also shape the future direction of AI research by highlighting areas for innovation and enhancement [16]. Utilizing pre-trained language models (e.g., BERT, GPT) to evaluate the coherence and fluency of generated text. This can provide insights into the quality of AI-generated content. Language Model Evaluation serves to understand the strengths and weaknesses of different models. Through comprehensive benchmarking and analysis, helping to validate the effectiveness of the models in capturing the nuances of human language and contributes to advancing the state-of-the-art in generative AI technology.

### 3.8. Diversity Metrics

Diversity metrics play a crucial role in assessing the breadth and depth of language utilization within the generated essays. By examining the range of words and expressions employed, these metrics provide insights into the linguistic diversity present in the AI-generated content compared to human-authored texts. Through the application of various diversity metrics, such as lexical diversity measures or diversity indices, the extent to which the language used in the generated essays reflects a wide array of vocabulary and linguistic forms can be quantified. Exploring diversity metrics to measure the variety of words and expressions used in the essays highlight differences in language richness. Prioritizing diversity alongside accuracy is fundamental [17]. Natural language generation systems aspire to do more than simply generate accurate outputs; they aim to create responses that exhibit diversity and nuance.

### 3.9. Topic Modeling

Topic modeling tasks involve identifying groups of words (topics) within a corpus of text, a task that proves challenging to accomplish manually due to the vastness of data. Many topic modeling techniques have been created to automatically pull out topics from short texts. These include non-negative matrix factorization, random projection, principal component analysis, latent semantic analysis, and Latent Dirichlet allocation (LDA). Evaluating the performance of these methods based on metrics such as topic quality, recall, precision, F-score, and topic coherence reveals that latent Dirichlet allocation yields meaningful extracted topics and achieves favorable results [18]. Discovering themes within essays through techniques such as Latent Dirichlet Allocation (LDA) serves as a pivotal aspect of this research.

$$\rho(\omega|topic) = \frac{n_{topic,\omega} + \beta}{\sum_{\omega'} + \beta}$$

Where $n_{topic,\omega}$ is the count of word ω assigned to topic $topic$ and β a smoothing parameter. By employing topic modeling methods like LDA, latent topics present in the text is uncovered, shedding light on thematic distinctions between AI-generated and human-authored writing.

### 3.10. N-gram Analysis

Examining the frequency and distribution of n-grams (sequences of adjacent words) is a fundamental aspect of this research. By delving into n-gram analysis, deeper insights into language patterns and styles present in the text is gained, utilizing n-grams for sentiment analysis at the article level offers numerous advantages, as longer phrases tend to convey less ambiguity in terms of their polarity. Employing a discriminating classifier alongside high-order n-grams as features has demonstrated comparable, if not superior, sentiment analysis performance compared to state-of-the-art methods on large-scale datasets [19]. Incorporating n-gram features serves as a solution for scenarios where traditional feature extraction methods may fall short in capturing nuanced language patterns and subtle variations in sentiment expression. This can provide insights into language patterns and styles.

### 3.11. Semantic Analysis

Exploring the semantics of words and phrases is crucial for discerning nuanced differences between AI-generated and human-authored content. Semantic analysis constitutes a fundamental aspect of NLP approaches, offering insights into the contextual meaning of sentences and paragraphs. Semantic analysis involves scrutinizing the essence and context of language, shedding light on the intricate interplay between linguistic elements [20]. By deciphering the semantic significance of vocabulary, deeper insights into the underlying subject matter and themes conveyed within the text can be gained.

## 4. LINGUISTIC ANALYSIS

To explore the richness and diversity of vocabulary in both AI-generated and human-authored texts where: Total number of AI-generated essays is 37232, Total number of human-authored essays is 62768.

Various metrics are utilized for this assessment, including the calculation of unique tokens, determination of average token length, and the application of tools like the Type-Token Ratio (TTR). The analysis reveals intriguing insights into the vocabulary size of both AI-generated and human-authored texts. Specifically, the AI-generated corpus comprises 66,088 unique tokens, while the human-authored counterpart contains 85,529 unique tokens.

## 4.1. Vocabulary Exploration

Exploring vocabulary richness, word clouds are utilized to visualize word frequency in both AI-generated and human-authored essays in Fig 4 and 5. Word clouds present a visual depiction of frequently occurring words, with larger font sizes indicating higher frequencies. This approach allows insights into predominant themes and prevalent vocabulary across each corpus. In the case of AI-generated essays, all essays are aggregated into a unified string and a word cloud is generated to depict common words. Similarly, for human-authored essays, the same procedure is followed. These word clouds offer a rapid overview of vocabulary distribution, aiding in the identification of pivotal terms and recurring themes.

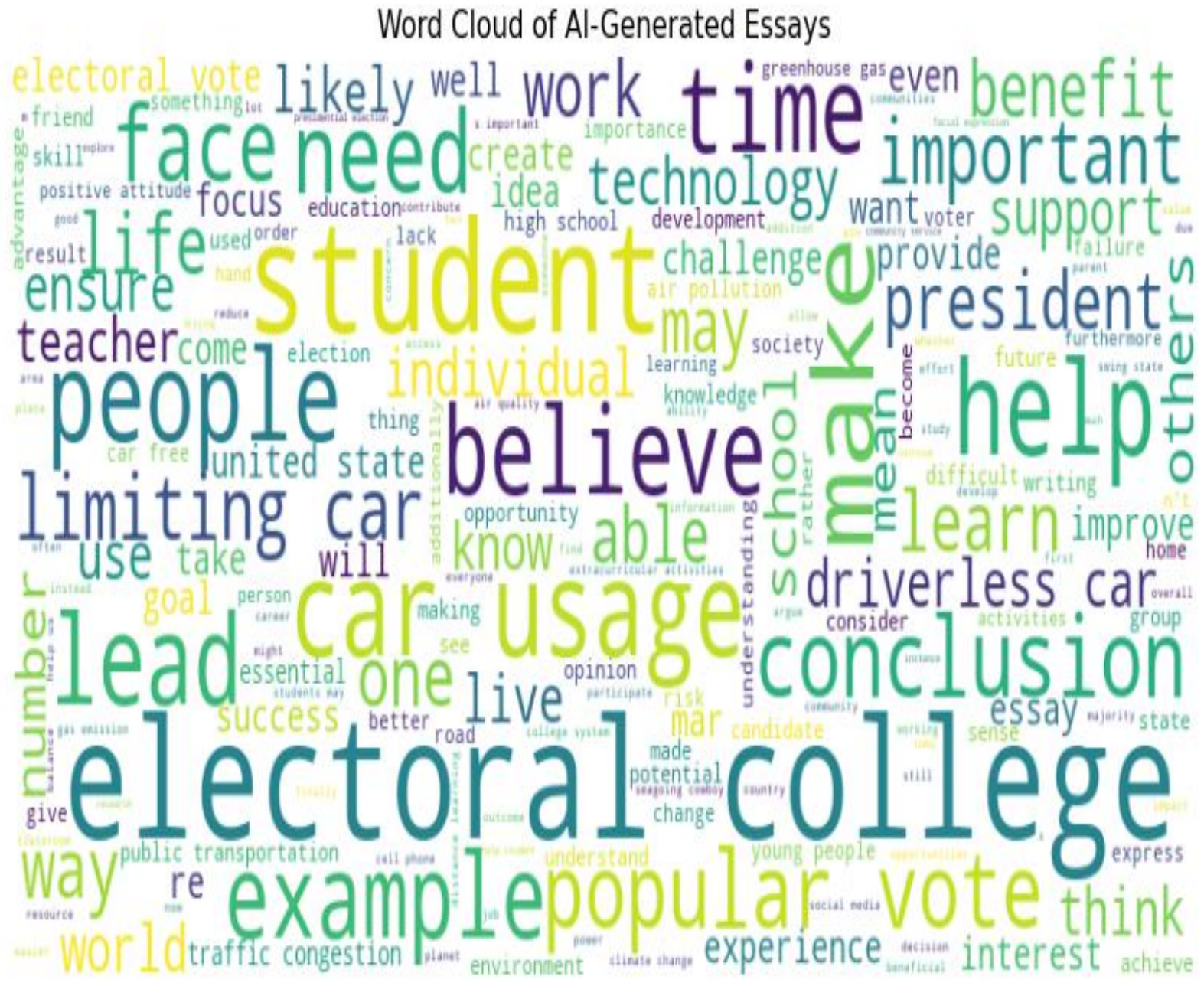

Fig 4. Word Cloud of AI-Generated Essays

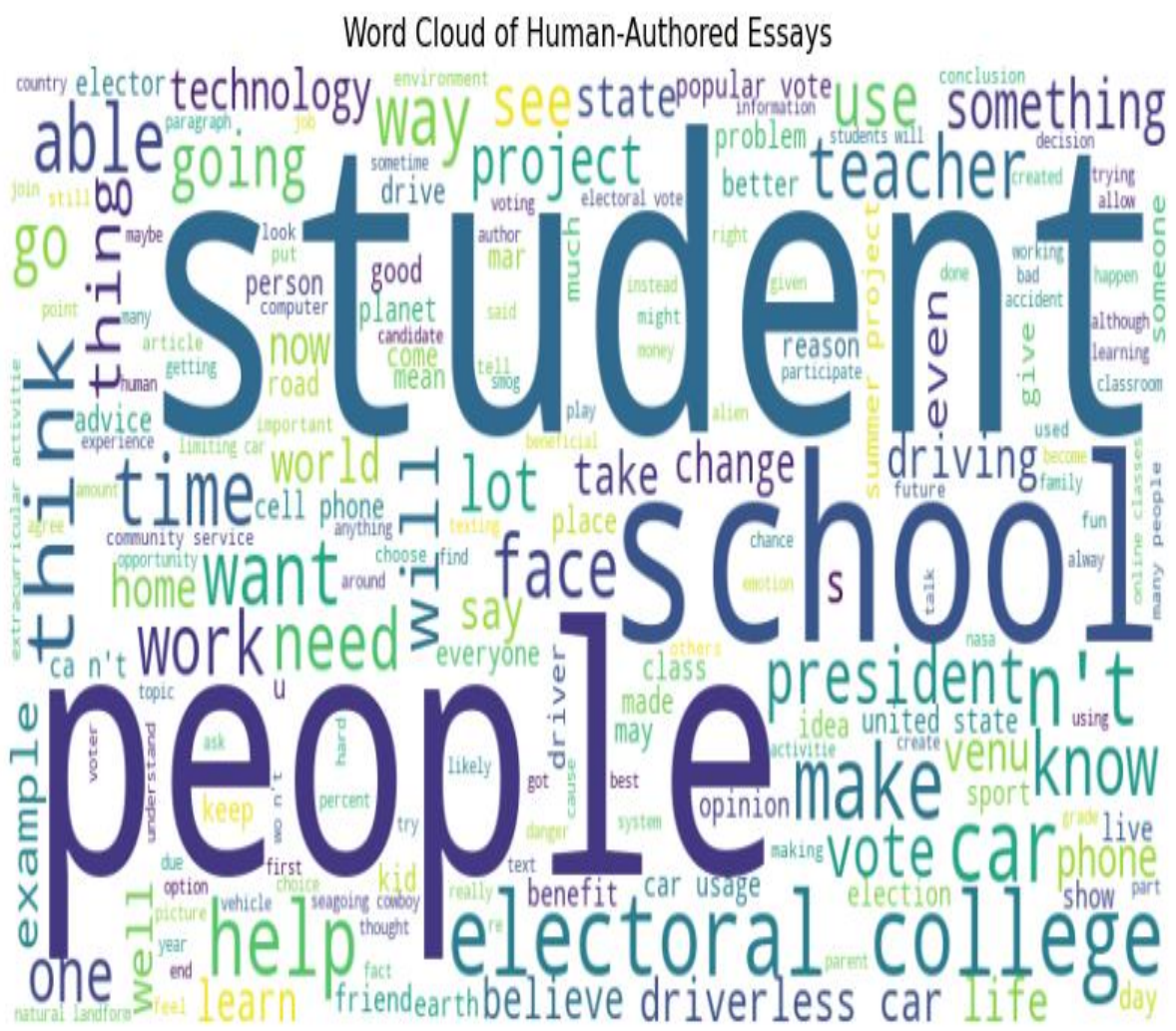

Fig 5. Word Cloud of Human-Authored Essays

## 4.3. Comparative Metrics

To quantify the differences between AI-generated and human-authored texts, we use cosine similarity, Jaccard similarity, or other similarity measures which will provide a nuanced understanding of the degree of similarity or dissimilarity between the two sets of texts. This analysis serves to highlight the extent to which AI-generated content aligns with or diverges from human-authored content in terms of linguistic patterns, vocabulary usage, and overall textual characteristics. Visualizing the word frequency distributions, we employed bar plots to illustrate the most common words in both sets of essays. This graphical representation in Fig 6 allows for a comparative assessment of vocabulary usage between AI-generated and human-authored texts, offering a glimpse into the linguistic characteristics of each corpus. Analysis showed that: Average AI-generated Sentence Length is 390.9946687384467 and the Average Human-authored Sentence Length is 468.8100457932443.

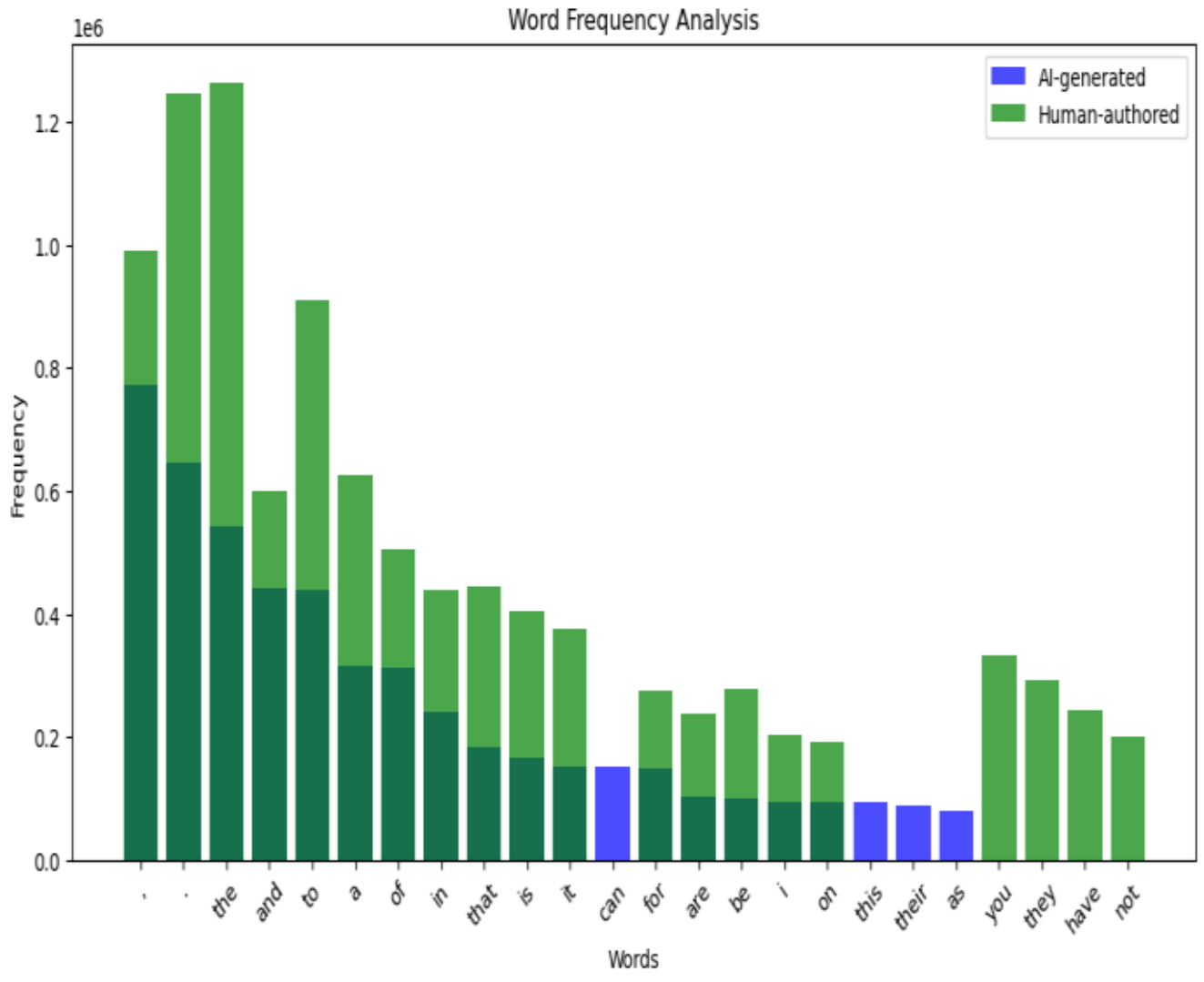

Fig 6. Word Frequency analysis of AI-generated and Human Authored essays

## 4.4. Part-of-Speech Tagging

To perform POS tagging on both AI-generated and human-authored texts for further linguistic analysis, we utilize the Natural Language Toolkit (NLTK), we tag each word in the texts with its corresponding part-of-speech category, such as noun, verb, adjective, etc. Following the POS tagging, we calculate the distribution of POS tags in each set of texts. This allows us to ascertain the frequency of occurrence of different parts of speech and gain insights into the syntactic structure and grammatical patterns prevalent in AI-generated and human-authored content.

## 4.5. Creativity Metrics

The analysis of creativity metrics in our research on AI-generated and human-authored essays reveals several key insights into the differences and similarities in creative writing styles between the two groups as shown on Table 1.

Table 2. Average Creativity Metrics for AI-generated and Human-authored Essays

| Metric | AI-Generated Essays | Human-Authored Essays |
|---|---|---|
| Total Words | 343 | 421 |
| Vocabulary Diversity | 157 | 174 |
| Average Word Length | 4.97 | 4.39 |
| Fluency (Score) | 0.99 | 1.0 |
| Novelty (Score) | 0.47 | 0.44 |

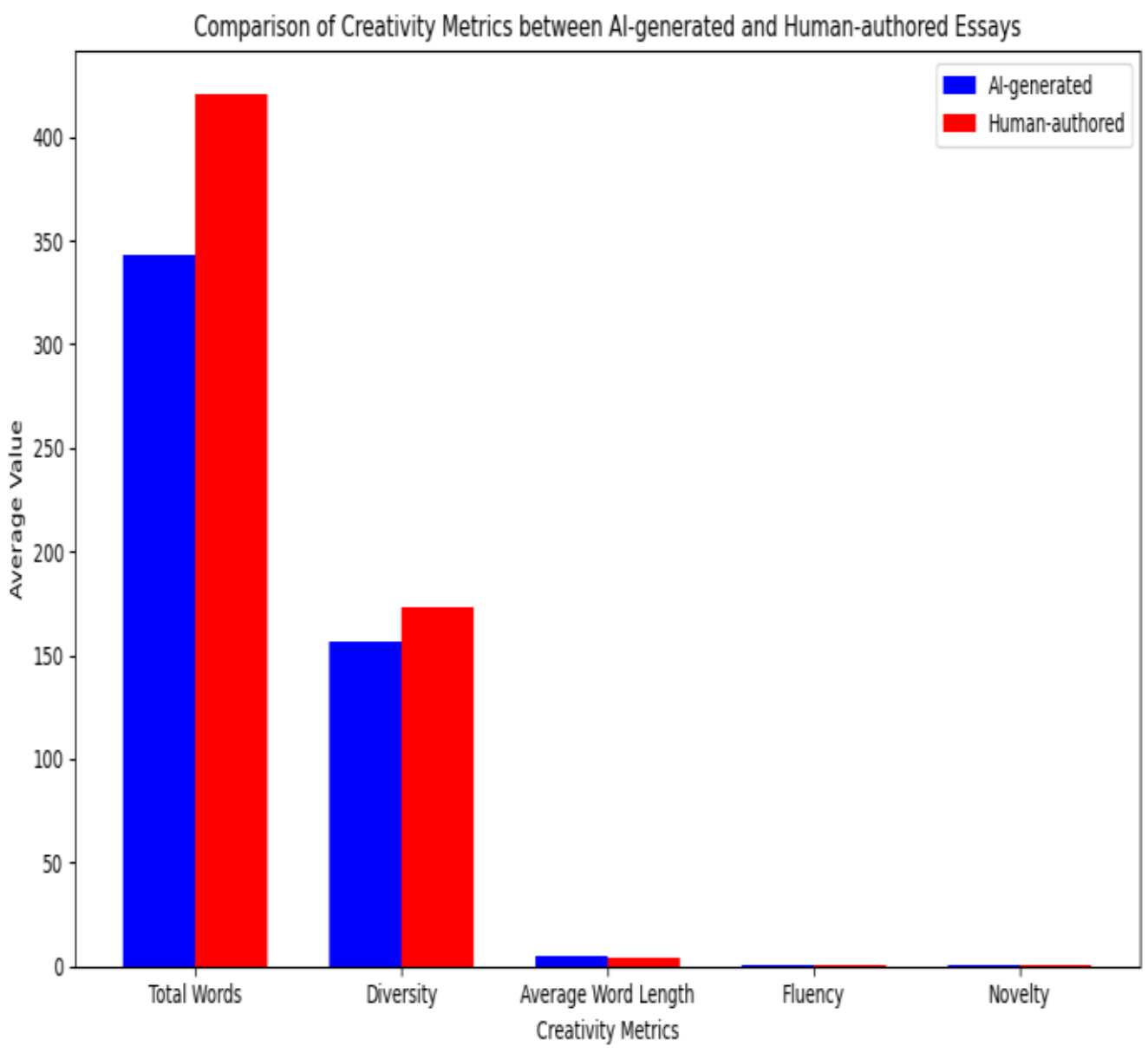

Fig.7. Comparison of Creativity Metrics between AI-generated and Human-authored Essays

The comparative analysis between AI-generated and human-authored essays reveals nuanced differences in several creativity metrics as shown in Fig 7. Human-authored essays tend to be longer on average, with approximately 421 words compared to 343 words in AI-generated essays. Vocabulary diversity is higher in human-authored texts, which contain around 174 unique words per essay, whereas AI-generated essays have about 157 unique words. The average word length is slightly longer in AI-generated essays (4.97 characters) compared to human-authored ones (4.39 characters). Both types of essays exhibit high fluency, indicating smooth and coherent writing. Interestingly, AI-generated essays have a marginally higher novelty score (0.47) compared to human-authored essays (0.44), the novelty score provides insights into the diversity of vocabulary and the extent to which the text introduces new or uncommon language constructs suggesting a potential for generating more original content through AI systems. Overall, these findings highlight the longer, more diverse vocabulary of human-authored essays and the slightly higher novelty in AI-generated essays, providing insights into the creative writing styles of both groups

## 4.6. Bias Analysis

The average gender bias in AI-generated essays was found to be approximately 0.86, while in human-authored essays, it was approximately 1.73. This suggests a slightly lower prevalence of gender bias in AI-generated essays compared to human-authored essays.

## 4.7. Biased Topic Presence

A significant number of essays, 5047 from AI-generated and 2781 from human-authored, were found to contain biased topics such as race, gender, religion, ethnicity, sexuality, and disability. This indicates that biased topics are present in a considerable portion of both AI-generated and human-authored essays. The average topic bias per AI-generated essay is higher (13.56%) compared to human-authored essays (4.43%). This indicates a higher prevalence of biased topics in AI-generated essays than in human-authored essays.

## 4.8. Sentiment Analysis

Sentiment analysis was conducted to analyze the distribution of sentiment polarity scores in AI-generated and human-authored essays. The sentiment polarity scores ranged from -0.2 to 0.4, with a peak around 0.1, indicating a predominantly positive sentiment in Fig 8. Both AI-generated and human-authored essays lean positive. The sentiment distribution shows a peak around 0.1 for both categories, indicating a generally positive slant. While the overall distribution is similar, human-authored essays appear to have slightly more positive sentiment with a small bump towards the positive side of the scale (0.2). The peak frequency for human-authored essays being slightly higher than that for AI-generated essays, suggests a stronger positive sentiment trend in human-authored content. Sentiment analysis hereby highlights a predominantly positive sentiment in both AI-generated and human-authored essays, with human-authored essays exhibiting slightly stronger positive sentiment trends

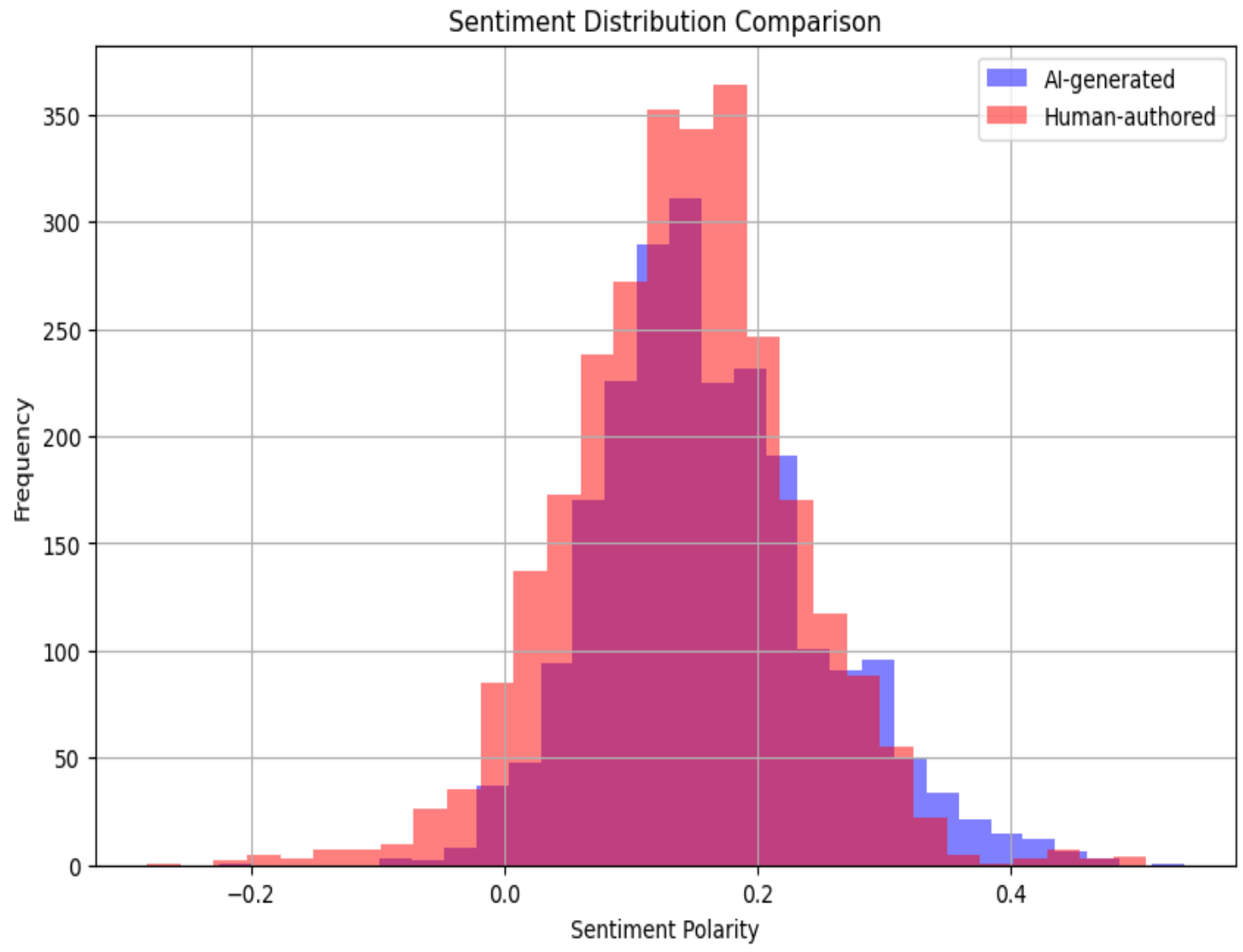

Fig.8. Sentiment distribution comparison between the AI-generated and Human authored essays

## 5. FEATURE ENGINEERING

Feature engineering involved several transformations and calculations on the original dataset as shown in Fig 9, to extract meaningful attributes where the dataset consisted of text samples along with a binary indicator column denoting whether the text was generated by an AI system orauthored by a human.

The length of each essay was calculated by splitting the text into words and counting the number of words.

| | text | generated | essay_length |
|---|---|---|---|
| **131513** | phones and driving do n't mix well thousands o... | 0 | 711 |
| **459347** | the power of positivity it is widely believed ... | 1 | 466 |
| **396563** | is there such thing that a computer can define... | 0 | 404 |
| **62286** | dear teacher_name , i think it would n't be fa... | 0 | 205 |
| **192144** | dear senator of florida . i am here today too ... | 0 | 900 |

Fig.9. Random Subset Preview Table including essay length

The transformations on our dataset resulted in the creation of several engineered features, including **sentence count, paragraph count, average word length, coherence, originality, complexity, engagement, sentiment, named entity counts, gender pronoun counts**, and indicators for **cultural references**. In the process of preparing the data for analysis, several important steps were taken to ensure the dataset was appropriately structured and ready for use in classification tasks. One such step involved the utilization of one-hot encoding techniques on the 'named_entities' and 'cultural_references' columns. This method was employed to transform categorical variables into binary features, facilitating their incorporation into machine learning models. Specifically, each distinct named entity (such as locations, organizations, and persons) and cultural reference (such as art, history, and literature) present in the essays was assigned a separate binary column, indicating its presence or absence in each sample. Coherence scores were computed using the TextBlob library to measure the logical flow and connectivity of ideas within each essay by analyzing sentiment and subjectivity. Originality was quantified by calculating the ratio of unique words to total words, assessing the uniqueness of vocabulary and novelty of ideas. Linguistic complexity was evaluated by considering word diversity and sentence structure, using average word length and sentence count as proxies. Engagement levels were determined by analyzing sentiment polarity to measure the emotional impact on the reader. Named entities, such as locations, organizations, and persons, were identified using a function for named entity recognition. Gender pronoun occurrences were counted using a function for gender pronoun analysis. Additionally, cultural references, including mentions of art, history, and literature, were detected using a function designed for identifying such references.

# 6. METHODOLOGY

## 6.1. Random Forest Classifier Model and Random Forest Algorithm

The Random Forest algorithm is a popular ensemble learning technique used for classification and regression tasks. It belongs to the family of decision tree algorithms and works by constructing multiple decision trees during the training phase and outputting the class that is the mode of the classes (classification) or mean prediction (regression) of the individual trees. Additionally, the 'gender_pronouns' column, initially represented as dictionaries containing counts of male and female pronouns, was further processed to enhance its usability in the classification task. Specifically, the column was converted into separate columns for male and female pronoun counts, with the assumption that the dictionary keys were 'male' and 'female'.

### 6.1.1. Data Splitting

The first step in our classification approach involved splitting the dataset into training and testing sets to assess the model's performance. This division was crucial for evaluating the model's ability to generalize to unseen data. Using the train_test_split function from the sklearn.model_selection module, we allocated 80% of the data for training and reserved the remaining 20% for testing. This resulted in a training set comprising 80,000 samples and a test set containing 20,000 samples

### 6.1.2. Model Training

Following data splitting, we proceeded to train a Random Forest Classifier on the prepared dataset. Leveraging the ensemble learning technique provided by the Random Forest algorithm, we aimed to build a robust predictive model capable of capturing complex relationships within the data.

The classifier was initialized with default hyperparameters, and the fit method was employed to train it on the training data. This process enabled the model to learn from the features in the training set and their corresponding labels.

### 6.1.3. Model Evaluation

Once the classifier was trained, we evaluated its performance on the unseen test set using various performance metrics. The classifier's predictions were generated for the test set using the predict method, and the accuracy was computed by comparing these predictions with the true labels. The achieved accuracy of 0.9088 indicated that the model correctly classified approximately 90.88% of the test samples.

Table 3. Random Forest Model Performance Metrics for AI-generated and Human-authored Texts

| Class | 0 (AI-generated texts | Class 1 (Human-authored texts) |
|---|---|---|
| Precision | 0.91 | 0.91 |
| Recall | 0.95 | 0.84 |
| F1-score | 0.93 | 0.87 |

Additionally, Table 2 displays we computed a comprehensive classification report, which provided insights into the model's precision, recall, and F1-score for each class (AI-generated and human-authored texts). These metrics provide insights into the model's performance for each class. Class 0 (AI-generated texts) shows slightly higher recall compared to precision, indicating that the model correctly identifies a high proportion of AI-generated texts out of all actual AI-generated texts. Class 1 (Human-authored texts) demonstrates slightly lower recall than precision, suggesting that the model correctly identifies a slightly lower proportion of human-authored texts out of all actual human-authored texts. The F1-score, which is the harmonic mean of precision and recall, provides a balance between these two metrics, giving an overall measure of the model's performance for each class.

### 6.1.4. Model Result

To gain further insights into the factors influencing the classifier's decisions, we analyzed the feature importance scores provided by the trained Random Forest model. These scores, derived from the Gini impurity measure, indicated the relative importance of each feature in contributing to the model's predictive performance. The top features identified as shown in Table 3 included 'average_word_length', 'complexity', 'originality', and 'sentiment', suggesting that these linguistic and stylistic attributes played pivotal roles in distinguishing between AI-generated and human-authored texts.

Table 4. Feature Importance Scores in Random Forest Model

| Feature Number | Feature | Importance |
|---|---|---|
| 2 | average_word_length | 0.465880 |
| 5 | complexity | 0.150404 |
| 4 | originality | 0.143104 |
| 7 | sentiment | 0.118323 |
| 0 | sentence_count | 0.098538 |
| 12 | cultural_references_History | 0.014522 |
| 11 | cultural_references_Art | 0.007858 |
| 13 | cultural_references_Literature | 0.001370 |

The ROC curve and the area under the curve (AUC) were computed to evaluate the performance of the Random Forest classifier. The ROC curve in Fig 11 visually represents the trade-off between the true positive rate (sensitivity) and the false positive rate (1 - specificity) across different threshold values. The AUC quantifies the classifier's ability to distinguish between the positive and negative classes, with a higher AUC indicating better performance. By plotting the ROC curve and calculating the AUC, we gained valuable insights into how well the classifier discriminates between AI-generated and human-authored texts.

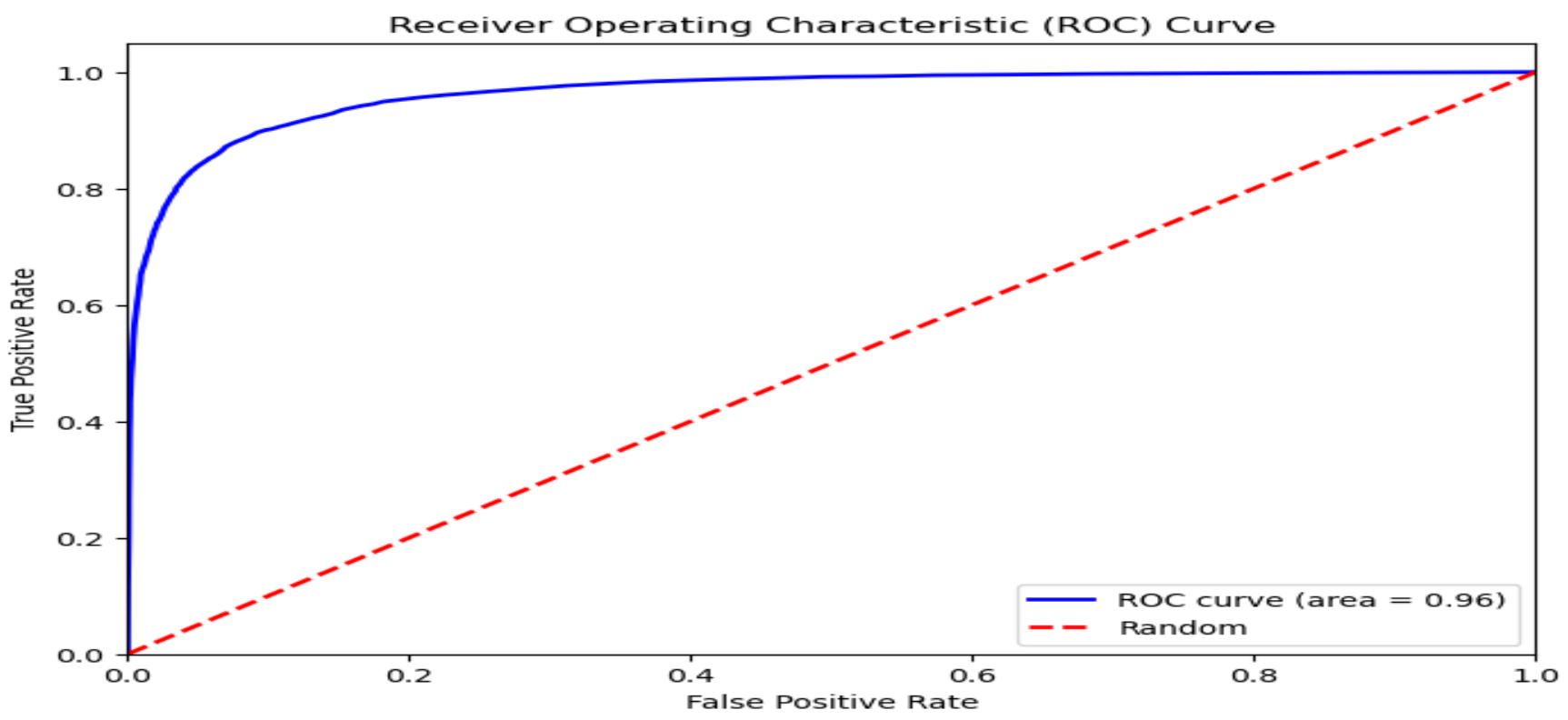

Fig.11. ROC Curve for the Random Forest Model

The area under the ROC curve (0.96) indicates high discriminatory power, suggesting that the model effectively separates the two classes. The interpretation of results provided valuable insights into the underlying mechanisms driving the classifier's decision-making process, enhancing our understanding of the dataset and the classification problem at hand.

### 6.2. BERT (Bidirectional Encoder Representations from Transformers)

Google's BERT is a potent pre-trained language representation model, by leveraging its bidirectional context understanding capabilities, BERT has the potential to capture intricate linguistic patterns and semantic relationships.

Unlike other models trained on unidirectional context, BERT learns from the entire input sequence simultaneously. Leveraging the pre-trained BERT architecture, the model was fine-tuned using the training subset in a supervised learning paradigm. During training, the model iteratively optimized its parameters to minimize a predefined loss function. The training process spanned three epochs: Epoch 1, Loss: 0.6613135295152665, Epoch 2, Loss: 0.6611969586849212 and Epoch 3, Loss: 0.6611056702375412F. The BERT model, employed for text classification, demonstrated an accuracy of approximately 63.02%. Despite its sophisticated bidirectional context understanding capabilities, BERT's performance was lower than anticipated in this study, likely due to the complexity of the dataset. This result highlights the challenges in accurately classifying textual data using deep learning models and displays the need for further optimization and exploration of model architectures for improved performance

## 7. LIMITATIONS

The selection of features for model training and the choice of evaluation metrics were deliberate decisions made to define the project's boundaries. Limitations of the study include the inherent ambiguity and subjectivity present in textual data, which posed significant challenges. These factors introduced noise and variability into the classification process, thereby limiting the project's precision. Assumptions pertain to underlying beliefs or conditions that guided decisions during data preprocessing and model implementation. The project assumed the availability of labeled training data and the representativeness of the dataset used. These assumptions influenced the project's methodology and the interpretation of its findings

## 8. SUMMARY

In this project, we investigated the intricacies of classifying textual data as either human- or AI-authored. By leveraging sophisticated methods in natural language processing, such as feature engineering and machine learning algorithms, we explored the complex patterns and linguistic properties present in written material. Our research yielded several significant findings that enhanced our understanding of this classification problem. Primarily, our analysis demonstrated the effectiveness of using conventional machine learning algorithms like the Random Forest Classifier to address this classification challenge. The Random Forest Classifier exhibited commendable accuracy, achieving approximately 91%. This highlights the robustness of traditional machine learning approaches in handling complex linguistic data. We also identified the key linguistic features that significantly influenced the categorization process through our investigation of feature importance. Features such as average word length, complexity, originality, and sentiment emerged as pivotal factors in distinguishing between AI-generated and human-authored texts. Understanding the relative importance of these features provided valuable insights into the underlying mechanisms driving the classifier's decision-making process.

However, our analysis also revealed inherent challenges in accurately classifying textual data, particularly when dealing with subtle linguistic nuances and stylistic variations.

Despite achieving notable accuracy rates, the Random Forest Classifier faced limitations in precisely discerning between AI-generated and human-authored texts. These results highlight the ongoing need for further investigation and improvement in the field of natural language processing, as we continue to grapple with the complexities of understanding and generating human language.

## 9. CONCLUSION

This project contributes to the evolving field of natural language processing by providing insights into the strengths and limitations of machine learning models in classifying text data. We explored the effectiveness of machine learning models, particularly the Random Forest Classifier, in distinguishing AI-generated texts from human-authored ones. Our findings pave the way for future advancements in this domain by highlighting the nuances in linguistic traits, creativity metrics, and biases inherent in AI-generated and human-authored texts. To improve classification accuracy and robustness, future research efforts could focus on enhancing existing models and developing cutting-edge methods. Investigating the integration of contextual data and domain-specific knowledge into the classification process may provide valuable insights for improving model performance. Additionally, exploring different datasets and incorporating more linguistic features could lead to a more thorough understanding of the subtleties involved in textual classification. Future work may also examine the role of advanced deep learning models and hybrid approaches combining traditional machine learning with deep learning techniques. Furthermore, research into the ethical implications and societal impact of AI-generated content is crucial to ensure responsible AI usage and adherence to cultural norms and sensitivities. Ultimately, by building upon the findings of this project and continuing to push the boundaries of natural language processing, we can strive towards more accurate and nuanced classification models in the realm of textual analysis. This ongoing exploration will contribute to the responsible advancement of AI technologies and their integration into various fields such as literature, communication, and knowledge dissemination.

## AUTHORS

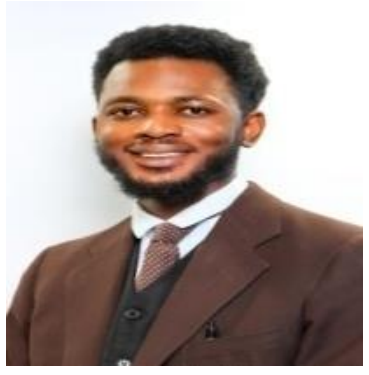

**Mayowa Jesse Akinwande** is a seasoned product management professional with nearly a decade of experience in financial services and operational technologies. He holds a Bachelor of Science (BSc) degree in Physics and Education from Universisty of Lagos, and a Master of Science (MSc) degree in Computer Science and Quantitative Methods from Austin Peay State University, specializing in Data Management and Analysis. With six years of experience in consumer lending risk, retail banking and product management at the United Bank for Africa, Mayowa has a deep understanding of financial product operations and IT. He is passionate about enhancing AI capabilities and has led various research initiatives on promoting financial inclusion, achieving operational excellence, and utilizing data-driven methodologies to deliver impactful products worldwide. During his time at Austin Peay, where he served as a Graduate Teaching Assistant, he founded and managed CampusSpark, a web-based platform which uses machine learning algorithms to connect students and their organizations with external brands for mentorship and sponsorships related to student innovations. Mayowa has authored several papers on Artificial Intelligence, ethics of AI, and credit risk, reflecting his dedication to advancing the field of AI and its responsible application.In his current role as a Digital Product Manager at American Express within the Digital Risk Products and Data Strategy team of the Credit and Fraud Business Unit, Mayowa continues to apply his knowledge and passion to innovate and excel in the ever-evolving fields of AI, digital product development, and data management technology.

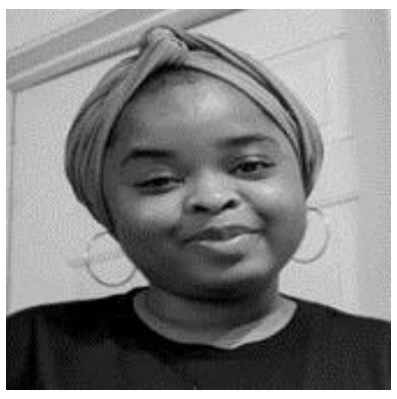

**Oluwaseyi Adeliyi** is a Product Manager at American Express working on Risk Data management for authorization and decisioning systems. In his previous role at COSMOS Research Center at the University of Arkansas at Little Rock, where he also obtained his Master's degree in Information Science; he led research in the areas of social computing and online coordinated actions, covering disinformation campaigns, crowd manipulation strategies, and inorganic activity detection on content-centric social media platforms such as YouTube. He also transitioned his research into a user-facing tool, VTracker, which provides numerous analytical capabilities on YouTube content, such as chronological metadata visualizations, sentiment & emotion analysis, network clustering, narrative analysis, and more. Oluwaseyi has published his work in several proceedings including established proceedings such as IEEE and ACM. He also has a book chapter in Springer's Cyber Security and Social Media Applications.

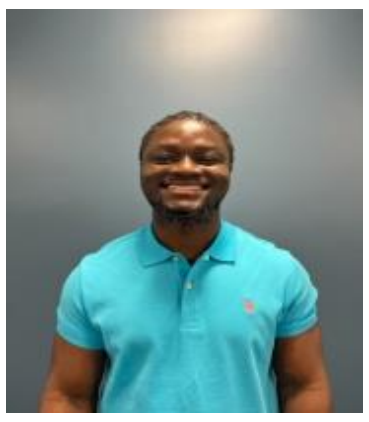

**Toyyibat T. Yussuph** holds a bachelor's degree in management and accounting (Nigeria) and a master's degree in management information systems (USA). Her extensive experience in product development, data strategy, and financial and data analysis has enabled her to transform and automate several critical insights into fraud trends and risk mitigation strategies as a Product Development Manager with American Express. She has a passion for innovation and leveraging big data to solve Credit and Fraud risks within the Financial Service industry